\documentclass[11pt]{article}
\usepackage{moriond07qcd_wessels,epsfig}

\newcommand{\Mall}{M_{\mbox{\footnotesize{all}}}}
\newcommand{\pmin}{p_{\mbox{\footnotesize{min}}}}
\newcommand{\Supt}{\sum P_T}
\newcommand{\Phat}{\hat{P}}

\newcommand{\ejnp}{\mbox{$e$-$j$-$\nu$}}

\newcommand{\mujnp}{\mbox{$\mu$-$j$-$\nu$}}

\newcommand{\jjjj}{\mbox{$j$-$j$-$j$-$j$}}

\newcommand{\jjjjnp}{\mbox{$j$-$j$-$j$-$j$-$\nu$}}

\begin{document}
\vspace*{4cm}
\title{\boldmath A GENERAL SEARCH FOR NEW PHENOMENA IN $E^-P$ SCATTERING\\ AT HERA}

\author{
  MARTIN WESSELS\\ 
  on behalf of the H1 Collaboration\\[0.1cm]}

\address{
  Deutsches Elektronen-Synchrotron DESY, Notkestra\ss{}e 85\\
  22607 Hamburg, Germany}

\maketitle\abstracts{
  A model-independent search for deviations from the Standard Model prediction 
  is performed in $e^- p$ collisions at HERA~II using H1 data recorded during the 
  years $2005$ and $2006$, corresponding to an integrated luminosity of 
  $159$ $\mbox{pb}^{-1}$. All event topologies involving isolated electrons, photons, 
  muons, neutrinos and jets with high transverse momenta are investigated in a single 
  analysis. Events are assigned to exclusive classes according to their final state.
  A statistical algorithm is used to search for deviations from the Standard Model in 
  the distributions of the scalar sum of transverse momenta or invariant mass of final 
  state particles and to quantify their significance.
  A good agreement with the Standard Model prediction is observed in most
  of the event classes. No significant deviation is found in the phase--space and 
  event topologies covered by this analysis.}

\section{Introduction}
\label{sec:intro}
At HERA electrons~\footnote{
  In this paper ``electrons'' refers to both electrons and positrons, if
  not otherwise stated.}
and protons collide at a centre-of-mass energy of up to $319$~GeV. 
These high-energy electron-proton interactions provide a testing ground for 
the Standard Model (SM) complementary to $e^+e^-$ and $p\overline{p}$ scattering.\\
The approach described in this paper~\cite{H1Coll:prel06-161} closely follows 
the strategy of the previously published H1 analysis using HERA I data~\cite{Aktas:2004pz}.
It consists of a comprehensive and generic search for deviations from the SM 
prediction at large transverse momenta. The analysis covers phase--space regions 
where the SM prediction is sufficiently precise to detect anomalies 
and does not rely on assumptions concerning the characteristics of any 
SM extension. Using the complete HERA II $e^-p$ data sample, this is the first 
general search performed on a large data set from electron-proton collisions.

\section{Data analysis}
\label{sec:analysis}
The event sample studied consists of the full $2005$--$2006$ HERA~II $e^-p$ 
data set, corresponding to an integrated luminosity of $159$~pb$^{-1}$.
All final states with at least two objects with $P_T>20$~GeV 
in the polar angle range $10^\circ<\theta<140^\circ$ are investigated. 
Considered objects are
electrons ($e$), photons ($\gamma$), muons ($\mu$), jets ($j$) and 
neutrinos ($\nu$) (or non-interacting particles). 
The identification criteria for each type of object are similar to those 
applied in the published HERA~I analysis, ensuring an 
unambiguous identification while keeping high efficiencies. 
All objects are required to be isolated from each other by a minimum 
distance $R$ of $1$ unit in the $\eta-\phi$ plane.
The events are classified into exclusive event classes according to the 
number and types of objects. 
This exclusive classification ensures a clear separation of the final states 
and allows an unambiguous statistical interpretation.\\
As this analysis investigates all final state topologies of $ep$ interactions
at high transverse momentum, a precise and reliable estimate of all relevant HERA 
processes is needed. Hence, several Monte Carlo generators are used to generate a 
large number of events in all event classes, carefully avoiding double-counting of 
processes. The simulation contains the order $\alpha_S$ matrix elements for QCD 
processes, while second order $\alpha$ matrix elements are used to calculate QED 
processes. Additional jets are modelled using leading logarithmic 
parton showers as representation of higher order QCD radiation.
All processes are generated with a luminosity significantly higher than that of 
the data.\\[0.2cm]
\begin{figure}[t]
  \center
  \includegraphics[width=\textwidth]{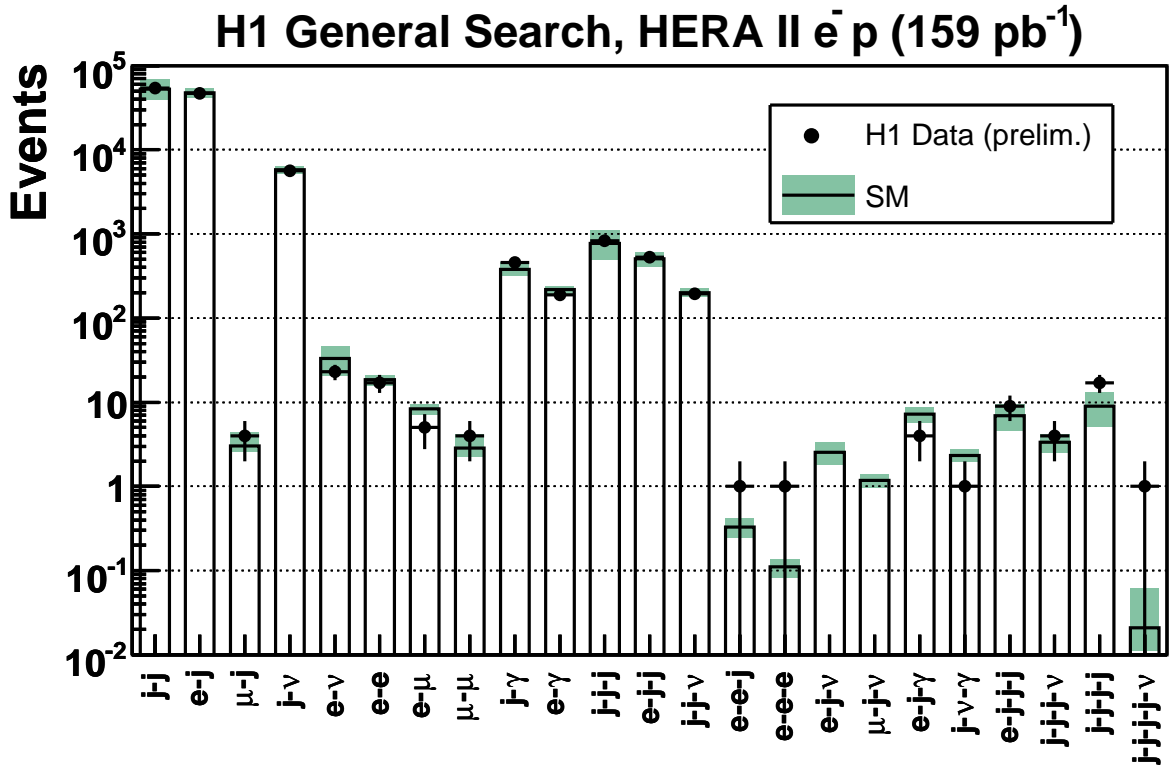}  
  \caption{The data and the SM expectation for all event classes 
    with observed data events or a SM expectation greater than $1$ event.
    The analysed data sample corresponds to an integrated luminosity of 
    $159$~pb$^{-1}$. The error bands on the predictions include model 
    uncertainties and experimental systematic errors added in quadrature.}
  \label{fig:summaryplot}
\end{figure}
The results of the data analysis are summarised in figure~\ref{fig:summaryplot},
which presents the event yields subdivided into event classes for the data and SM 
expectation. All event classes with observed data events or a SM prediction
greater than 1 event are shown~\footnote{
  The $\mu-\nu$ event class is discarded from the present analysis. It is
  dominated by events in which a poorly reconstructed muon gives rise to 
  missing transverse momentum, faking the neutrino signature.}.
In each class, a good agreement between the number of observed data events 
and the SM prediction is seen.\\
No data events are observed in the event classes \mujnp~and \ejnp~
where the largest discrepancy between the data and the SM prediction 
was found in the analysis of the HERA~I data (see section~\ref{sec:comparison}). 
Those classes correspond mainly to high $P_T$ $W$ production with 
subsequent leptonic decay, where deviations in the $e^+p$ data 
continue to be observed~\cite{H1Coll:prel07-063}. 
The total SM expectation amounts to $1.2 \pm 0.2$ and $2.5 \pm 0.8$ in 
the \mujnp~and \ejnp~classes, respectively.

\section{Search for deviations}
\label{sec:search}
In order to quantify the level of agreement between the data and SM expectation 
and to identify regions of possible deviations, the invariant mass $\Mall$
and sum of transverse momenta $\Supt$ distributions of all reliable event classes 
are systematically investigated using the same search algorithm as developed
for the previous publication. 
A region is defined as a sample of connected histogram bins, which have at least the 
size of twice the resolution of the observable. A statistical estimator $p$ is defined 
to determine the region of most interest by calculating the probability that the SM 
expectation fluctuates upwards or downwards to the data. This estimator is derived 
from the convolution of a Poisson probability density function (pdf) to account for 
statistical errors with a Gaussian pdf to include systematic uncertainties.
A possible sign of new physics is found if the expectation significantly disagrees 
with the data, and thus the region of most interest (greatest deviation) 
is given by the region having the smallest $p$-value, $\pmin$. This method finds 
narrow resonances and single outstanding events, as well as signals spread over 
large regions of phase--space in distributions of any shape.\\
The fact that somewhere in the distribution a fluctuation with a value $\pmin$
might occur is taken into account by calculating the probability $\Phat$ to observe 
a deviation with a $p$-value $\pmin$ at any position in the distribution.
Thus $\Phat$ is the central measure of significance of the deviation found. 
\begin{figure}[t]
  \begin{center}
    \epsfig{figure=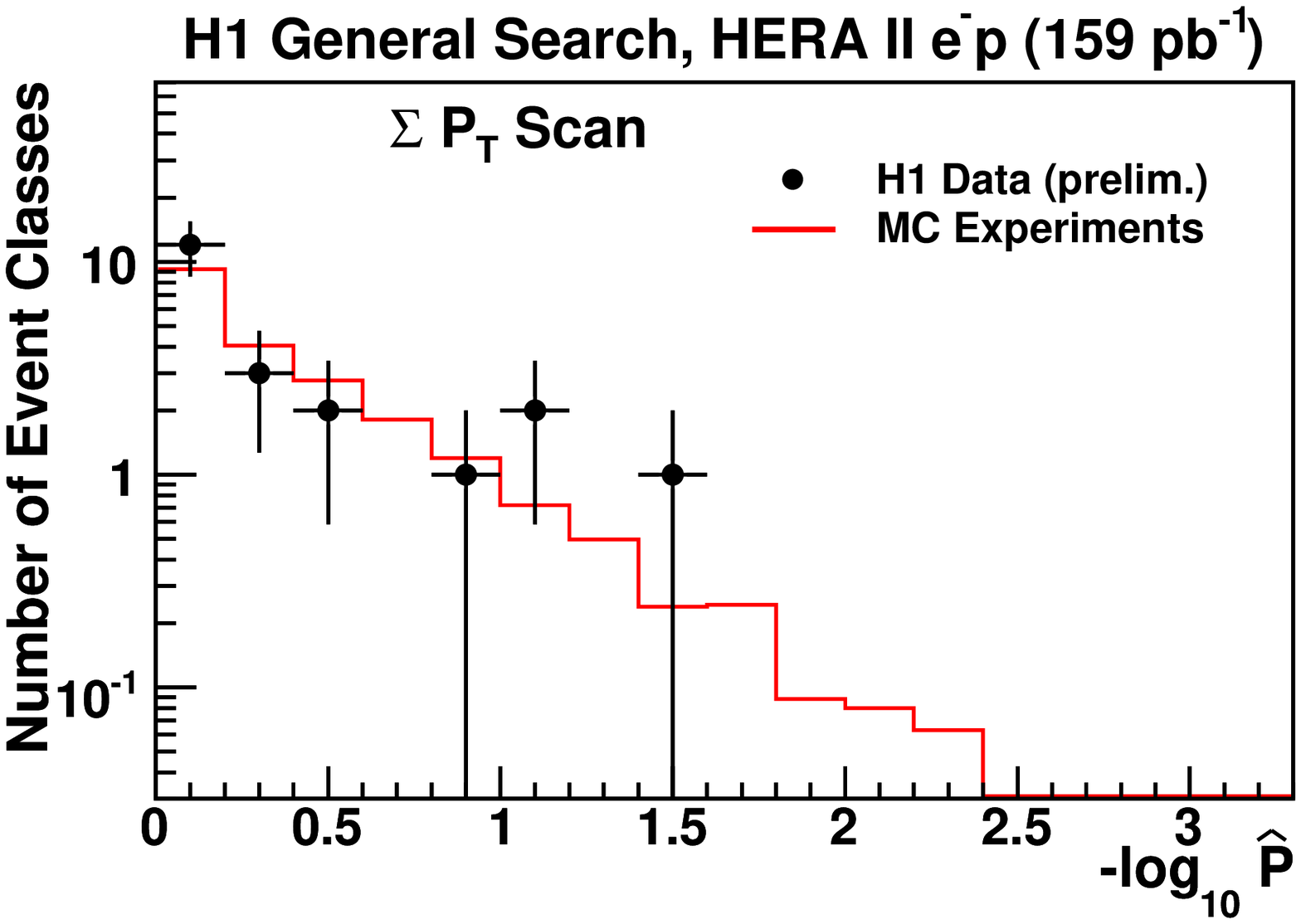,width=0.5\textwidth}\hfill
    \epsfig{figure=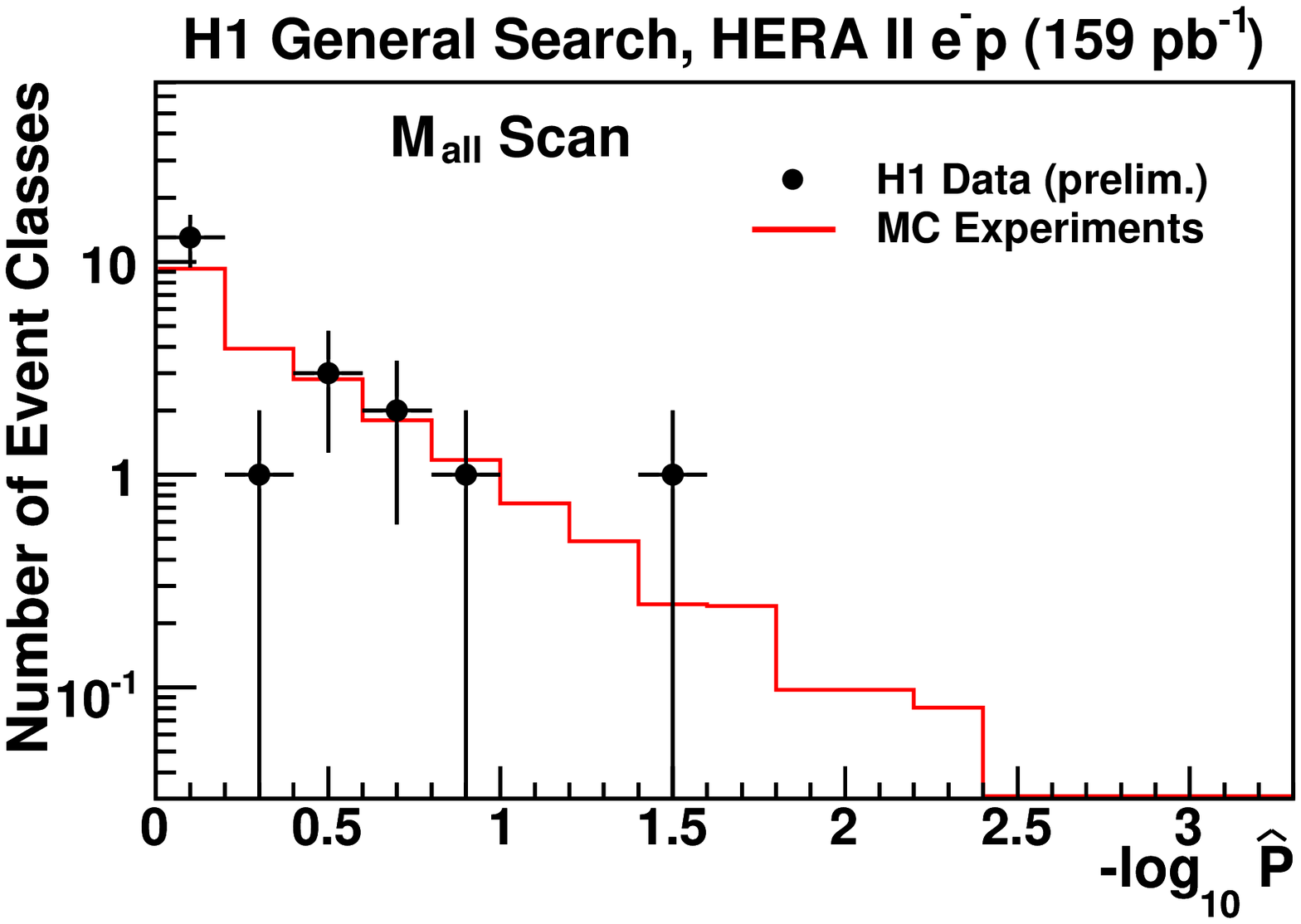,width=0.5\textwidth}
    \caption{The $-\log{\Phat}$ values for the data event classes and the 
      expected distribution from MC experiments as derived by investigating
      the $\Supt$ distributions (left) and $\Mall$ distributions
      (right) with the search algorithm.}
    \label{fig:scan}
  \end{center}
\end{figure}
To determine $\Phat$, hypothetical data histograms are diced according to the 
pdf of the expectation. The value of $\Phat$ is then defined as the fraction of 
hypothetical data histograms with a $\pmin$-value smaller than the $\pmin$-value 
obtained from the data, and consequently the event class of most interest for a search 
is the one with the smallest $\Phat$-value.\\
The overall level of agreement between the data and SM expectation can be quantified
further by taking into account the large number of event classes studied in this
analysis. Among all classes, there is some chance that small $\Phat$ values occur.
This probability can be calculated by replacing all data distributions by hypothetical 
Monte Carlo (MC) distributions based on the SM expectation. The complete statistical 
algorithm is applied on this MC experiment, representing a single HERA experiment 
with an integrated luminosity of $159$~pb$^{-1}$. The expectation for the $\Phat$ values 
of the data is then given by the distribution of $\Phat_{SM}$ values derived from
many MC experiments. In the case that deviations arise from
statistical or systematical fluctuations only, the distribution of $\Phat$-values 
obtained from data and MC experiments are compatible.\\[0.2cm]
The results of the search for deviations between data and SM expectation
are summarised in figure~\ref{fig:scan}. Shown is the negative base--$10$ logarithm 
of the $\Phat$ values obtained from the real data compared to the expectation
derived from a large set of MC experiments. The comparison is presented separately 
for the scans of the $\Mall$ and $\Supt$ distributions. All $\Phat$ values range from
$0.01$ to $0.99$, corresponding to event classes where no significant 
discrepancy between data and SM expectation is observed. These results are in good
agreement with the expectation from MC experiments.\\[0.2cm]
Although data events are observed in the \jjjj~and \jjjjnp~event classes,
no reliable $\Phat$ values can be calculated for these classes
due to uncertainties of the SM prediction at highest $\Mall$ and $\Supt$
values~\cite{Aktas:2004pz}. These event classes are not considered 
to search for deviations from the SM in this extreme kinematic domain. 
Consequently, these event classes are not taken into account to determine 
the overall degree of agreement between the data and the SM.

\section{Comparison with HERA~I results}
\label{sec:comparison}
While good agreement in all event classes is observed between the data and the 
SM prediction in the present HERA~II analysis, some discrepancy is found in the 
previously published general search on the HERA~I data.
Complementary to the pure $e^-p$ event sample studied here, the HERA~I data 
set is largely dominated by positron-proton collisions. 
There the most significant deviation is found in the \mujnp~event class
with $\Phat$ values of $0.01$ and $0.001$ for the scan of the $\Mall$ and 
$\Supt$ distributions, respectively. The global probability to find at least
one event class with a $\Phat$ value smaller than that of the \mujnp~class 
in the HERA~I data amounts to 28\% for the $\Mall$ and 3\% 
for the $\Supt$ distributions.

\section{Conclusions}
\label{sec:conclusions}
The data collected with the H1 experiment during the years $2005$--$2006$ 
(HERA~II) have been investigated for deviations from the SM prediction at 
high transverse momentum. All event topologies involving isolated electrons, 
photons, muons, neutrinos and jets are investigated in a single analysis.
This is the first general search performed on a large set of data from 
electron--proton collisions. A good agreement between data and SM expectation 
is found in most event classes. In each event class the invariant mass and sum 
of transverse momenta distributions of particles have been systematically 
searched for deviations using a statistical algorithm. 
No significant deviation from the SM is observed in the phase--space and event 
topologies covered by this analysis.

\end{document}